# Reply to: "Comment on (t, n) Threshold d-level Quantum Secret Sharing"

Chuang Li[1] •Longwei Zhang[2] •Xiuli Song[2]

*(1. School of computer science and technology, Chongqing university of posts and telecommunications, Chongqing 400065, China*

*2. School of cyber security and information law, Chongqing university of posts and telecommunications, Chongqing 400065, China;)*

**Abstract:** A corresponding comment, raised by Kao and Hwang, claims that the reconstructor $Bob_1$ is unable to obtain the expected secret information in (*t, n*) Threshold d-level Quantum Secret Sharing (TDQSS) [Scientific Reports, Vol. 7, No. 1 (2017), pp.6366]. In this reply, we show the TDQSS scheme can obtain the dealer's secret information in the condition of adding a step on disentanglement.

In Ref. [1], the authors proposed a (*t, n*) threshold quantum secret sharing (TDQSS) scheme with *d*-dimensional quantum states. It can resist some common attacks with lower computation and communication costs than other similar QSS schemes. In Ref. [2], Kao and Hwang claimed that the reconstructor $Bob_1$ is unable to obtain the dealer's secret information in the TDQSS scheme of Ref. [1]. In fact, we show the TDQSS scheme can obtain the dealer's secret information in the condition of adding a step on disentanglement. We will explain it as follows.

In TDQSS scheme, between Step 6 and Step 7, a step on disentanglement is added as follow:

**Addition step:** After all participants have performed generalized Pauli operators, $Bob_1$ performs respectively *d*-dimensional CNOT operation on the $Bob_r$'s particle for $r = 2,...,t-1$, where $Bob_1$'s particle is the control qudit and $Bob_r$'s particle $r = 2,...,t-1$ is the target qudit. In Eq. (12) of TDQSS, the state $|\varphi_3\rangle$ is evolves as

$$|\varphi_4\rangle = \frac{1}{\sqrt{d}} \sum_{k=0}^{d-1} \omega^{(\sum_{r=0}^{t} s_r) \cdot k} \left( CNOT(|k\rangle_1, |k\rangle_2) \otimes \cdots \otimes \left( CNOT(|k\rangle_1, |k\rangle_t) \right) \right)$$
$$= \frac{1}{\sqrt{d}} \sum_{k=0}^{d-1} \omega^{(\sum_{r=0}^{t} s_r) \cdot k} |k\rangle_1 |0\rangle_2 |0\rangle_3 \cdots |0\rangle_t \quad (1)$$

Then in the step 7 of TDQSS, when $Bob_1$ applies $QFT^{-1}$ to his own particle $|k\rangle_1$, the state $|\varphi_4\rangle$ is evolves as

$$|\varphi_5\rangle = QFT^{-1}\left(\frac{1}{\sqrt{d}} \sum_{k=0}^{d-1} \omega^{(\sum_{r=0}^{t} s_r) \cdot k} |k\rangle_1\right) = \sum_{r=0}^{t} s_r \quad (2)$$

After the above operation, $Bob_1$ can obtain the secret information through measures his particle in the computational basis.

We can draw a conclusion from the above steps, TDQSS scheme can correctly recover the secret information in the condition of adding a step on disentanglement.

**Example:** An example is given to illustrate the correctness of our view, when $d=4, t=3$, and $a_0 = 3$, in Step 6 of TDQSS scheme, the state $|\varphi_3\rangle$ is

$$|\varphi_3\rangle = \frac{1}{\sqrt{4}} \sum_{k=0}^{3} \omega^{3 \cdot k} |k\rangle_1 |k\rangle_2 |k\rangle_3 \quad (3)$$

Then, after the addition step is performed state $|\varphi_3\rangle$ should evolve as $|\varphi_4\rangle$

$$|\varphi_4\rangle = \frac{1}{\sqrt{4}} \sum_{k=0}^{3} \omega^{3 \cdot k} |k\rangle_1 |0\rangle_2 |0\rangle_3 \quad (4)$$

From Eq. (4), it can be seen that the entanglement among the three particles is released. We only focus on the Bob$_1$'s particles. When the $QFT^{-1}$ is performed on Bob$_1$'s particles, we obtains $QFT^{-1}(\frac{1}{\sqrt{4}} \sum_{k=0}^{3} \omega^{3 \cdot k} |k\rangle) = |3\rangle$. The particle measurement results are equal to the original secret $f(0)$.

**Simulation**: In Step 6 of TDQSSS, each participant $Bob_r$ $(r=1,2,\ldots,t)$ performs the generalized Pauli operator $U_{0,s_r}$ according to his shadow $s_r$, and $U_{0,s_r}$ is defined by

$$U_{0,s_r} = \sum_{k=0}^{d-1} \omega^{s_r \cdot k} |k\rangle_{rr}\langle k| \quad . \quad (5)$$

The d-dimensional $\omega^{s_r \cdot x}$ is rewritten as binary representation:

$$\omega^{s_r \cdot x} = \omega^{s_r \sum_{k=0}^{n-1} x_k \cdot 2^k} = \prod_{k=0}^{n-1} \omega^{s_r \cdot x_k \cdot 2^k} \quad (6)$$

where $\omega = e^{\frac{2\pi i}{d}}$. After the d-dimensional state $|x\rangle$ is performed Pauli operator $U_{0,s_r}$, it is evolves as

$$\begin{aligned} U_{0,s_r} |x\rangle &= \omega^{s_r \cdot x} |x\rangle = \left( \prod_{k=0}^{n-1} \omega^{s_r \cdot 2^k x_k} \right) |x_{n-1} \ldots x_0\rangle \\ &= \otimes_{k=0}^{n-1} \left( \omega^{s_r \cdot 2^k x_k} |x_k\rangle \right) \\ &= \otimes_{k=0}^{n-1} P_k |x_k\rangle \end{aligned} \quad (7)$$

where $P_k = \begin{bmatrix} 1 & 0 \\ 0 & e^{i\theta} \end{bmatrix}, \theta = 2\pi i \frac{s_r \cdot 2^k}{d}$.

We assume there are 3 participants $\{Bob_r | r=1,2,3\}$ and $Bob_1$ is a trusted reconstructor, The parameters are set as $d=4, t=3, a_0=3, s_1=0, s_2=1 s_3=2$. Because IBM Experiment only supports 2-dimension quantum space, in order to expand into 4-dimension quantum space, each participant owns two qubits. In Figure 1, $Bob_1$ holds qubits $q_0$ and $q_1$; $Bob_2$ holds qubits $q_2$ and $q_3$; $Bob_3$ holds qubits $q_4$ and $q_5$.

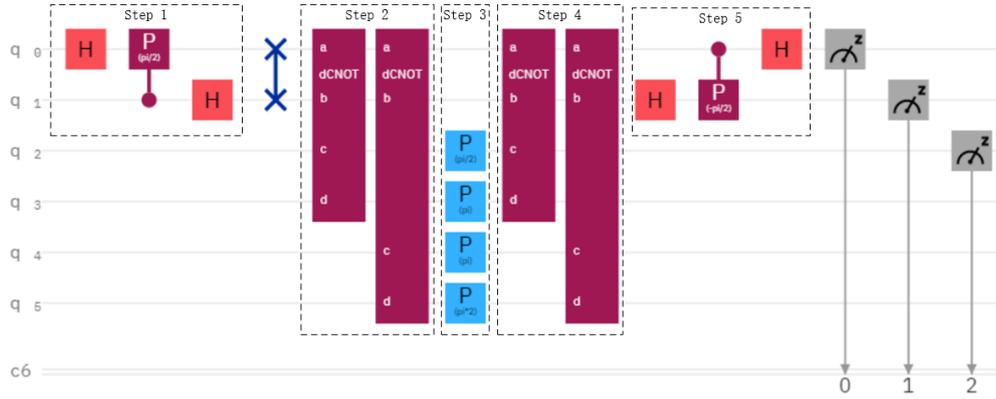

Figure 1 d-dimensional QSS Quantum Circuit

Step 1. Bob₁ performs QFT on his particle, where the gate phase $\theta_1 = \frac{\pi}{2}$.

Step 2. Bob₁ performs respectively *d*-dimensional CNOT operation *dCNOT* on the Bob₂'s particle and Bob₃'s particle. That is $CNOT(q_0, q_2)$, $CNOT(q_1, q_3)$, $CNOT(q_0, q_4)$, $CNOT(q_1, q_5)$, where $q_0, q_1$ are control particles, $q_2, q_3, q_4, q_5$ are target particles.

Step 3. According to the participants' shadows, Bob₁ performs Pauli operator $U_{0,0}$ on $q_0, q_1$, Bob₂ performs Pauli operator $U_{0,2}$ on $q_2, q_3$, Bob₃ performs Pauli operator $U_{0,1}$ on $q_4, q_5$. That is $P(\frac{\pi}{2}, q_2), P(\pi, q_3), P(\pi, q_4), P(2\pi, q_5)$.

Step 4. Bob₁ performs respectively *d*-dimensional CNOT operation on the Bob₂'s particle and Bob₃'s particle,

Step 5. Bob₁ performs the $QFT^{-1}$ (Inverse Quantum Fourier Transform) on his particle and measures his particle to obtain origin secret, where the gate phase $\theta_2 = -\frac{\pi}{2}$.

**Result verification:** Bob₁'s measurement results shown in the Figure 2.

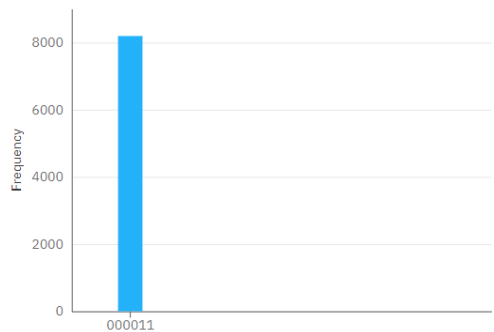

Figure 2 Measurement result

As is shown in figure 2, the experiment measurement output is $|000011\rangle$, Bob$_1$'s particle is $|11\rangle$, Bob$_2$'s particle is $|00\rangle$, and Bob$_3$'s particle is $|00\rangle$. The secret value $f(0)$ that needs to be reconstructed is 3, the binary form of 3 is 11. Bob$_1$'s particle measurement result is $f(0)' = 3$, which is equal to the original secret $f(0)$. Therefore, the experimental results are correct.

To summarize, we verify experimental the correctness of TDQSS in the condition of adding a step on disentanglement on IBM quantum cloud platform.